\documentclass[letterpaper]{article}
\usepackage[draft]{aaai2027}
\usepackage{times}
\usepackage{helvet}
\usepackage{courier}
\usepackage[hyphens]{url}
\usepackage{graphicx}
\usepackage{natbib}
\usepackage{caption}
\usepackage{booktabs}
\usepackage{multirow}
\usepackage{tabularx}
\usepackage[table]{xcolor}
\usepackage{amsmath}
\usepackage{amssymb}
\definecolor{semanticgray}{gray}{0.92}
\newcolumntype{Y}{>{\raggedleft\arraybackslash}X}
\title{VITAL-RAG: Invariance Race for Context Allocation in Coding Agents}
\author{
Zijian Lu\textsuperscript{\rm 1},
Yonghua Lu\textsuperscript{\rm 2},
Mingcai Chen\textsuperscript{\rm 1},
Yiping Zuo\textsuperscript{\rm 1},
Xin He\textsuperscript{\rm 1},
Weijun Wang\textsuperscript{\rm 3},
Weibei Fan\textsuperscript{\rm 1}
}
\affiliations{
\textsuperscript{\rm 1}Nanjing University of Posts and Telecommunications, Nanjing, China\\
\textsuperscript{\rm 2}Tianjin University, Tianjin, China\\
\textsuperscript{\rm 3}Institute for AI Industry Research, Tsinghua University, Beijing, China\\
18818732360@163.com, 3024202111@tju.edu.cn, chenmc@njupt.edu.cn\\
zuoyiping@njupt.edu.cn, xhe@njupt.edu.cn, wangweijun@air.tsinghua.edu.cn, wbfan@njupt.edu.cn
}

\begin{document}
\maketitle
\begin{abstract}
Coding agents often retrieve code from an entire repository, but only limited evidence can fit into the final model input. Conventional retrieval-augmented generation (RAG) for coding agents treats fragments from the same code object as separate results, so redundant views can occupy multiple context positions and crowd out useful code. Grouping fragments by code object reduces this redundancy, but can discard local information needed for the task. We describe this tension as an \emph{invariance race}: allocation should stay stable under redundant renderings but change when a fragment adds task-relevant semantics. To address this race, we introduce VITAL-RAG, which organizes evidence by canonical code object, keeps one query-relevant companion only when it adds semantics not already represented, and renders selected evidence under per-object and global token budgets. On RepoBench, VITAL-RAG improves Recall@4K from 39.59\% to 63.67\% while reducing evidence tokens by 35.63\%. Across three model backends, it matches or outperforms recent baselines on RepoClassBench and achieves the highest raw Pass@1 on RepoExec.
\end{abstract}

\section{Introduction}
\label{sec:introduction}
Many coding-agent tasks require more repository evidence than can fit in the final model input. Agents that generate code, repair bugs, modify interfaces, write tests, or maintain repositories must select among callees, types, tests, imports, documentation, and failure paths \citep{jimenez2024swebench,wang2025openhands,soni2026coding}. This creates a general context-allocation problem. A coding agent must decide which evidence receives enough of the bounded token budget to influence generation.

\begin{figure}[t]
\centering
\includegraphics[width=\columnwidth]{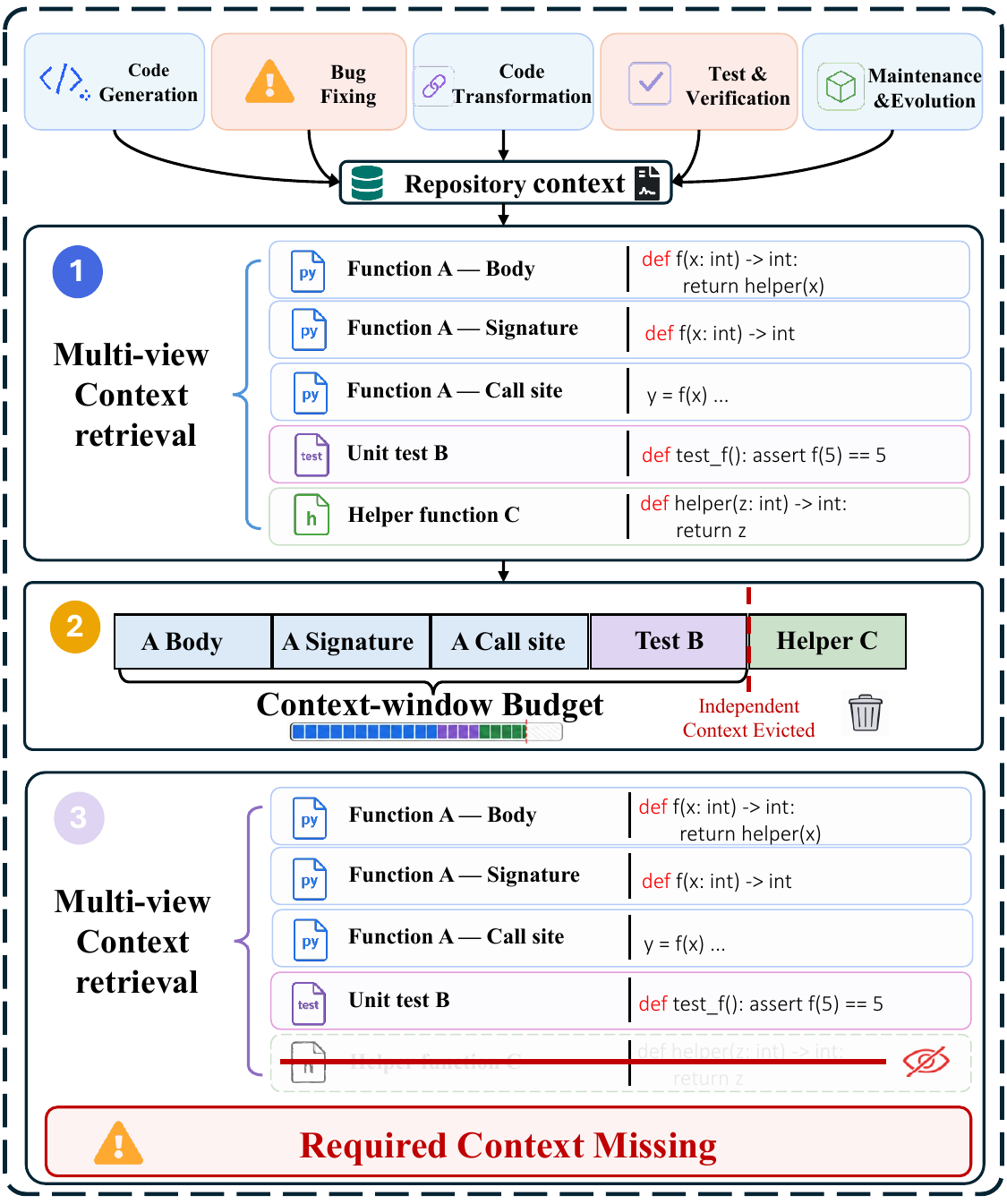}
\caption{A repository-context failure after successful retrieval. Repeated views of Function A consume slots, so required Helper C is clipped by the input budget.}
\label{fig:motivation}
\end{figure}

The problem is especially pronounced at repository scale. Existing work improves candidate discovery through selective retrieval and reranking \citep{wu2024repoformer,zhang2025coderag}, filters low-utility chunks \citep{huo2026reposhapley}, or enriches multi-view context \citep{liu2026reposcope,oh2026late}. However, a function body, signature, docstring, and call site may all describe the same code object. To our knowledge, prior work has not isolated and quantified the failure that occurs after correct evidence is retrieved but before generation, when repeated views compete with independent objects under a bounded rendering budget. We refer to this problem as \emph{repository evidence allocation for coding agents}. Figure~\ref{fig:motivation} presents the resulting retrieval-to-context failure.

Retrieval and allocation govern different stages of repository context construction. Selective retrieval and utility-aware filtering determine which candidates enter the pool \citep{wu2024repoformer,zhang2025coderag,huo2026reposhapley}, while long-context studies show that position affects model use \citep{liu2024lost}. Our focus is the intervening stage, where retrieved candidates are rendered into a bounded model input. As Figure~\ref{fig:motivation} shows, repeated views can occupy several positions even when the ranking is correct, leaving less context for independent evidence. Repository-scale agents therefore require an allocation rule that preserves broad discovery without allowing representation count to determine context share.

This allocation must operate at two coupled scales. Fragment-level selection is too fine-grained because multiple renderings of one code object receive separate opportunities to influence generation. Selection over only the globally strongest canonical code objects is too coarse-grained because a task-relevant validator, exception helper, or downstream call may be a distinct object within the same source region. We call this tension the \emph{invariance race}: allocation should remain invariant to redundant views but sensitive to task-relevant new information. We call the corresponding design principle \emph{selective invariance}. Context authority thus has a discrete component that selects objects and a continuous component that determines how much of each selected object survives the token budget.

We introduce VITAL-RAG, a repository context allocation layer placed between a provenance-preserving retriever and a coding-agent generator. The name expands to \emph{View-Invariant, Task-Aware Allocation Layer for RAG} and reflects its allocation principle. VITAL-RAG groups multi-view fragments into canonical code objects, keeps one query-relevant companion from the same source region, and assigns bounded token authority through compact evidence rendering. An optional program-semantic transfer extension arbitrates between programs produced from alternative portfolios. This design separates candidate discovery, object allocation, token allocation, and generation.

We make three contributions.

First, we identify and analyze a systematic retrieval-to-context failure in repository-level coding agents. Relevant evidence may be retrieved successfully but lost during bounded context construction. We trace this loss to authority multiplication and formalize the resulting invariance race between suppressing redundant views and preserving task-relevant local evidence.

Second, we introduce VITAL-RAG, a retriever-agnostic context allocation layer built on selective invariance. It groups multi-view fragments by canonical code object, keeps at most one query-relevant companion from the same source region, and controls token use through per-object and global budgets.

Third, we evaluate VITAL-RAG from evidence preservation to downstream code generation. On RepoBench, it raises Recall@4K from 39.59\% to 63.67\% while using 35.63\% fewer evidence tokens. It remains competitive on RepoClassBench and achieves the highest raw Pass@1 on RepoExec across three model backends.

\section{Related Work}
\label{sec:related-work}
\subsection{Repository-Level Retrieval for Coding Agents}
Repository-level code generation depends on cross-file evidence, as established by RepoBench and CrossCodeEval \citep{liu2024repobench,ding2023crosscodeeval}. Recent systems decide when to retrieve, follow dataflow or repository graphs, identify necessary knowledge, combine textual and structural signals, and preserve semantics across chunks \citep{wu2024repoformer,cheng2024dataflow,ouyang2025repograph,zhang2025coderag,shi2026aircoder,oh2026late}. SWE-bench, RepoClassBench, OpenHands, and OpenHands-Versa connect these advances to issue resolution, class generation, and tool-using agents \citep{jimenez2024swebench,deshpande2024repoclassbench,wang2025openhands,soni2026coding}. This broader evidence pool makes bounded context construction a distinct stage after retrieval.

\subsection{Evidence Selection and Context Allocation}
RAG selection determines which external information conditions generation \citep{lewis2020rag}. MMR and submodular objectives balance relevance with coverage, while coding-agent systems add selective retrieval, structural reranking, and learned filtering \citep{carbonell1998mmr,lin2011submodular,wu2024repoformer,cheng2024dataflow,zhang2025coderag,shi2026aircoder}. RepoShapley further models interaction-aware chunk utility \citep{huo2026reposhapley}. These methods move beyond raw recall but usually let candidate texts compete independently. Exact deduplication misses multiple renderings of one object, whereas coarse grouping can hide a query-relevant companion. The unresolved question is how code objects receive context authority before rendering.

\subsection{Context Robustness in Coding Agents}
Retrieved chunks can complement one another, interfere with decoding, lose semantics during chunking, or become less usable because of context position \citep{huo2026reposhapley,oh2026late,liu2024lost,yan2025rpo}. In coding agents, these choices affect repository edits, tests, and builds \citep{jimenez2024swebench,wang2025openhands,soni2026coding}. Repository pipelines sharpen the problem because parsers, graph indexes, and sliding windows often render one code object several ways. Context allocation must therefore ignore redundant views while retaining distinct nearby evidence, which motivates measuring evidence survival under a bounded model input.

\section{Empirical Evidence of Retrieval-to-Context Loss}
\label{sec:evidence-survival}
The literature distinguishes retrieval quality from context usefulness, but does not reveal how much already-retrieved evidence is lost during model input construction. We test this boundary before model-specific generation or execution. RepoBench provides labeled cross-file evidence at scale, letting us observe whether evidence found by retrieval remains available under a finite token budget. A fixed pre-method selection setting isolates this transition at the bounded model input.

\subsection{Dataset and Measurement Protocol}
RepoBench supplies repository-level completion tasks with labeled cross-file evidence \citep{liu2024repobench}. The source has 16,755 rows. Exact-row deduplication leaves 16,741, and leakage and invalid-label screening yields 16,490 eligible tasks from 2,798 repositories. The shared evaluation set contains 8,556 Java and 7,934 Python tasks.

We test whether required evidence enters the portfolio but its labeled content disappears from the bounded model input. Entity-First 4+1 fixes the discrete allocation to at most four canonical code objects and one ungated companion from a selected provenance fiber. Ranking, surface views, and token budget remain fixed, while exact text identity is tracked separately from provenance identity. Each object is capped at 1,024 tokens. Recall@5 measures portfolio entry, while Recall@4K grants credit only when the complete labeled snippet survives within 4,096 tokens. A path or clipped signature therefore does not count as preserved evidence. $\Delta\mathrm{Recall}$ is the difference between the two measures. Auth.@5 counts selected canonical code objects, and Tokens reports mean rendered length.

\subsection{Cross-Language Evidence Survival}
\begin{table}[t]
\centering
\small
\setlength{\tabcolsep}{1.0pt}
\begin{tabular*}{\columnwidth}{@{\extracolsep{\fill}}lrrrrrr@{}}
\toprule
\textbf{Scope} & \textbf{Tasks} & \textbf{R@5} & \textbf{Auth.@5} & \textbf{R@4K} & \textbf{$\Delta$Recall} & \textbf{Tokens} \\
\midrule
All & 16,490 & 64.16 & 4.422 & 39.59 & 24.57 & 2,720 \\
Java & 8,556 & 63.97 & 4.413 & 37.60 & 26.37 & 2,798 \\
Python & 7,934 & 64.37 & 4.432 & 41.73 & 22.64 & 2,637 \\
\bottomrule
\end{tabular*}
\caption{RepoBench evidence-survival gap under Entity-First 4+1.}
\label{tab:repobench-survival-gap}
\end{table}

Table~\ref{tab:repobench-survival-gap} shows $\Delta\mathrm{Recall}=24.57$ points over all 16,490 tasks: required evidence appears in the selected top five for 64.16\% of tasks, but survives in the 4K model input for only 39.59\%. The pattern is not confined to one language. $\Delta\mathrm{Recall}$ is 26.37 points on Java and 22.64 points on Python, despite similar candidate recall and authority counts. Thus, a substantial part of the failure occurs after candidate selection, when retrieved evidence competes for bounded input context.

The scale and consistency of $\Delta\mathrm{Recall}$ matter more than the absolute recall of this setting. The finding spans thousands of repository clusters and two languages with different syntax, type systems, and file organization. Java has the larger $\Delta\mathrm{Recall}$, but Python still loses more than one third of the evidence counted as retrieved at five. Because the selection setting and metric definitions are fixed across the two splits, the result is difficult to explain as a language-specific ranking artifact. It instead points to a shared stage of the pipeline: converting ranked repository candidates into a bounded model input.

This observation changes what a method must improve. Raising Recall@5 alone is insufficient when selected evidence can disappear during rendering. Conversely, a context constructor can improve Recall@4K without changing the candidate set if it assigns the budget more carefully. Having established this gap, we next trace the allocation mechanism that produces it.

\section{Analysis of Evidence Allocation}
\label{sec:failure-analysis}
The large-scale finding establishes that candidate recall and input-level recall can diverge substantially. We now analyze where this divergence comes from and why familiar context controls do not resolve it.

\subsection{Authority Multiplication from Surface Views}
A repository retriever can expose a signature, body slice, call site, or test fragment for the same code object. Let $q$ be the task, $v$ a retrieved view, $g_{\mathcal I}(v)$ the object assigned by repository index $\mathcal I$, and $B$ the evidence budget. When allocation treats each view independently, one object can claim several input positions.

The resulting context share is $\mathrm{share}(e)=\sum_{v:g_{\mathcal I}(v)=e}\mathrm{tokens}(v)$. We call the repeated opportunities created by multiple views of $e$ \emph{authority multiplication}. Object identity follows provenance rather than lexical similarity. Different strings from one code object share an authority claim, while similar strings from independent helpers remain distinct. Under budget $B$, repeated views can displace independent evidence even when the ranking is correct.

For example, a top-five list may contain the signature, body, and call-site view of method $a$, followed by dependencies $b$ and $c$. Surface allocation gives $a$ three positions. Exact deduplication cannot remove these distinct strings, and lexical diversity may retain them because they emphasize different tokens. If the required evidence is the full body of $c$, Recall@5 records success when $c$ enters the list, while earlier views of $a$ can leave only a clipped form of $c$ in the model input. The failure comes from counting one object's relevance several times before assigning the bounded budget.

\subsection{Limitations of Existing Allocation Controls}
\begin{table}[t]
\centering
\small
\setlength{\tabcolsep}{1.0pt}
\begin{tabular*}{\columnwidth}{@{\extracolsep{\fill}}llrrrr@{}}
\toprule
\textbf{Control} & \textbf{Method} & \textbf{R@5} & \textbf{Auth.@5} & \textbf{R@4K} & \textbf{Tokens} \\
\midrule
\multirow{2}{*}{\textit{Identity}} & Raw Surface & 46.56 & 2.860 & 25.08 & 3,254 \\
 & Exact Dedup & 46.51 & 2.859 & 25.10 & 3,253 \\
\midrule
\multirow{4}{*}{\textit{Coverage}} & Canonical-5 & 64.10 & 4.422 & 39.47 & 2,734 \\
 & File Quota & 63.98 & 4.424 & 39.42 & 2,714 \\
 & MMR & 64.21 & 4.424 & 39.41 & 2,738 \\
 & Entity-First 4+1 & 64.16 & 4.422 & 39.59 & 2,720 \\
\bottomrule
\end{tabular*}
\caption{Standard allocation controls on the 16,490 RepoBench tasks.}
\label{tab:repobench-ordinary-fixes}
\end{table}

Table~\ref{tab:repobench-ordinary-fixes} separates several possible explanations. Exact Dedup changes Recall@4K by only 0.02 points, showing that byte-identical repetition is not the main issue. Canonical-5, File Quota, MMR~\citep{carbonell1998mmr}, and Entity-First 4+1 all raise Recall@5 to about 64\% and represent about 4.42 authorities among five candidates, yet Recall@4K remains between 39.41\% and 39.59\%. These controls improve retrieval diversity or object coverage, but they do not determine which query-relevant meaning should survive the token budget.

The controls act at distinct levels. Exact Dedup hashes normalized surface text. Canonical-5 keeps the strongest candidate for each normalized path-and-identifier key. File Quota first takes candidates from previously unseen normalized paths and fills remaining slots from deferred candidates. MMR balances retrieval or query overlap against Jaccard redundancy with equal weight. Entity-First 4+1 fixes four canonical code objects and allows one ungated candidate from the provenance fiber of a selected object. Their similar Recall@4K values are informative because they remove different kinds of repetition.

Together, these controls leave a more specific bottleneck. Identity, query-specific refinement, and rendering must be decided jointly: the context builder must decide which objects own external positions, which local distinctions matter, and how much text each selected object may contribute.

\subsection{Selective Invariance as an Allocation Principle}
Canonical quotienting prevents one code object from owning several external positions, but a purely global object ranking creates the opposite risk. Repository semantics are often distributed across provenance-local companions. A handler and its validator, a class and its exception helper, or an interface wrapper and its conversion routine are distinct canonical code objects whose relevance is coupled by the task. Ignoring this local structure preserves identity but can erase a necessary distinction.

Selective invariance resolves the invariance race at two scales. Allocation is invariant to view multiplicity at the object scale and sensitive to query-relevant additions at the provenance-fiber scale. Let $\mathcal{S}_q(\mathcal{V})\subseteq\mathcal{V}$ be the representative views assigned external positions from a candidate set $\mathcal{V}$. For canonical code object $e$, define its authority multiplicity as
\begin{equation}
\begin{aligned}
\mu_q(e\mid\mathcal{V})
&=\left|\left\{v\in\mathcal{S}_q(\mathcal{V})
\mid g_{\mathcal I}(v)=e\right\}\right|,\\
&\leq 1,\qquad e\in g_{\mathcal I}(\mathcal{V}).
\end{aligned}
\label{eq:selective-invariance}
\end{equation}
The guarantee concerns slot ownership rather than absolute rank stability under arbitrary score perturbations. Adding more renderings may replace an object's representative, but it cannot create another external authority for that object. Task sensitivity is introduced separately by allowing one distinct, query-relevant object from a selected provenance fiber. This separation turns the observed gap into a concrete allocation problem: preserve object-level invariance while keeping query-relevant local semantics.

\section{VITAL-RAG}
\label{sec:vitalrag}
\subsection{Method Overview}
\begin{figure*}[t]
\centering
\includegraphics[width=0.95\textwidth]{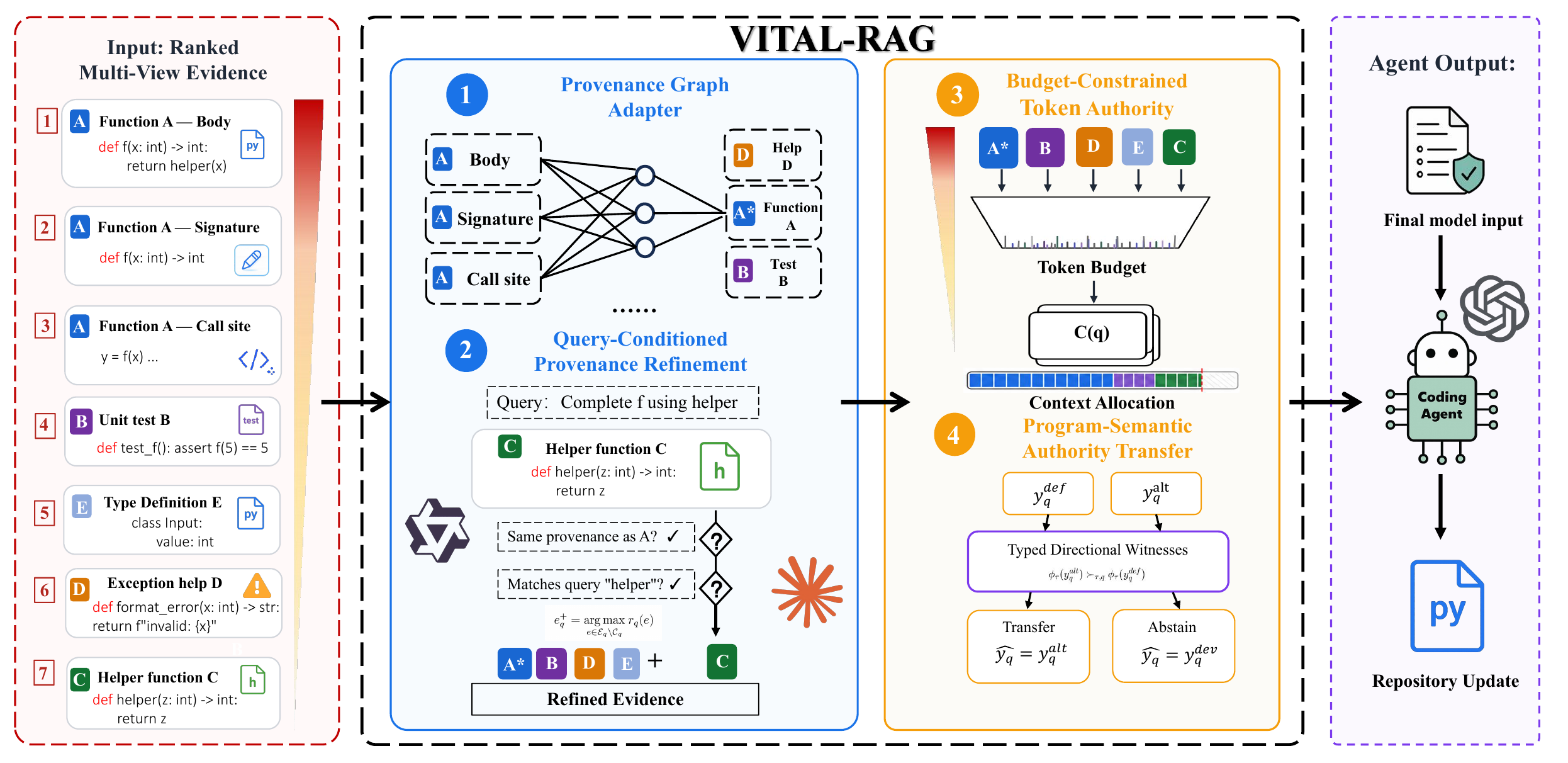}
\caption{VITAL-RAG's allocation path. A repository-index provenance adapter groups multi-view evidence by canonical code object. Query-conditioned refinement keeps one query-relevant companion, and budget-constrained rendering constructs $C(q)$. The lower-right transfer branch is an optional post-generation extension and does not alter context allocation.}
\label{fig:vitalrag}
\end{figure*}

VITAL-RAG is a repository context allocation layer between a provenance-preserving retriever and a coding-agent generator. It accepts ranked multi-view evidence from sparse, dense, or graph retrieval. Figure~\ref{fig:vitalrag} follows the allocation path used by the method. First, a repository-index provenance adapter groups body, signature, call-site, and related views by canonical code object and selects one representative for each object. Second, query-conditioned provenance refinement keeps one position for a distinct query-relevant companion. Third, budget-constrained token authority renders the resulting portfolio as $C(q)$, the model input. These three stages form the core allocation layer. The lower-right branch in Figure~\ref{fig:vitalrag} is an optional post-generation extension that arbitrates between two already generated programs and never changes $C(q)$. Formally, $q$ is a coding task, $\mathcal{V}_q$ is a finite set of retrieved views, $s_q:\mathcal{V}_q\rightarrow\mathbb{R}$ is the retrieval score, $K=5$ is the slot budget, and $B\in\mathbb{N}$ is the token budget.

\subsection{View-Quotiented Authority Allocation}
The first decision is which retrieved views make the same authority claim. VITAL-RAG obtains this relation from a provenance adapter supplied by repository index $\mathcal I$. The adapter $g_{\mathcal I}$ may use parser-qualified symbol IDs, graph nodes, or another stable object identifier. In our evaluation, the available index exposes a path $p(v)$ and identifier $n(v)$, so the adapter is instantiated as $g_{\mathcal I}(v)=(\operatorname{normpath}(p(v)),\operatorname{casefold}(n(v)))$. We write $\mathcal{E}_q=g_{\mathcal I}(\mathcal{V}_q)$ for the resulting object set and $\mathcal{V}_q(e)=\{v\in\mathcal{V}_q\mid g_{\mathcal I}(v)=e\}$ for the views of object $e$.

All views in $\mathcal{V}_q(e)$ now compete for one external position. Let $\succ_q$ denote their retrieval order. VITAL-RAG assigns the position to the strongest view and inherits its relevance score:
\begin{equation}
v_e^{\star}
    = \max_{\succ_q}\mathcal{V}_q(e),
\qquad
r_q(e)=s_q(v_e^{\star}).
\label{eq:object-authority}
\end{equation}
Equation~\ref{eq:object-authority} quotients authority without accumulating scores across repeated views. Low-overlap renderings of one object share a key, while similar helpers with different provenance remain separate. A new view may replace $v_e^{\star}$, but Eq.~\ref{eq:selective-invariance} prevents it from creating another position. Neither labels nor generator outputs enter this decision.

\subsection{Query-Conditioned Provenance Refinement}
Object-level invariance alone can be too rigid because a selected source region may contain a distinct helper needed by the task. VITAL-RAG therefore reserves one of the $K$ positions for task-aware refinement. The other $k_q=\min(K-1,|\mathcal{E}_q|)$ positions form the core set $\mathcal{C}_q=\operatorname{Top}_{k_q}(\mathcal{E}_q,r_q)$. With $K=5$, one reserved position is the smallest nonzero refinement that retains four globally ranked objects. The boundary ablation compares zero, one, and two companion positions. Let $\pi(e)$ denote an object's normalized source path, let $\pi(\mathcal{C}_q)$ denote the paths represented by the core, and let $T(x)$ extract identifier-like tokens from a task or view.

The reserved position is offered to the strongest non-core object that is both provenance-local and query-relevant:
\begin{equation}
\begin{aligned}
e_q^{+}
&= \underset{e\in\mathcal{E}_q\setminus\mathcal{C}_q}{\operatorname{arg\,max}}\ r_q(e)\\
\text{subject to}\quad
&\pi(e)\in\pi(\mathcal{C}_q),\quad
T(q)\cap T(v_e^{\star})\neq\varnothing.
\end{aligned}
\label{eq:task-aware-allocation}
\end{equation}
If the constraint set is nonempty, the final object portfolio is $\mathcal{A}_q=\mathcal{C}_q\cup\{e_q^{+}\}$. Otherwise, the highest-ranked remaining object fills the position, or the position is omitted when no object remains. Thus $|\mathcal{A}_q|=\min(K,|\mathcal{E}_q|)$, and every member owns one position. Crucially, $e_q^{+}$ is a distinct query-relevant companion rather than another fragment of a core object. The certificate uses only provenance and query tokens, never the labeled dependency.

\subsection{Budget-Constrained Token Authority}
Discrete allocation determines which objects own positions. Rendering determines their continuous token authority. Write the ordered allocation as $\mathcal{A}_q=(e_1,\ldots,e_m)$, where $m=|\mathcal{A}_q|$. Let $\bar\rho_q(e_i)$ be the unclipped descriptor and query-centered code window for object $e_i$, let $d_i=\ell(\bar\rho_q(e_i))$, and let $\rho_q(e_i,b_i)$ be its largest line-complete rendering of at most $b_i$ tokens.

The per-object cap $b_{\mathrm{obj}}$ prevents one object from consuming the model input. A conditional floor $\beta_q$ equals $b_{\min}$ when $B\ge mb_{\min}$ and zero otherwise. Let $R_i=B-\sum_{j<i}b_j$ be the unspent budget before object $e_i$. VITAL-RAG reserves the floor for every later object, then assigns the largest admissible share to the current object:
\begin{equation}
b_i
=\min\!\left\{d_i,b_{\mathrm{obj}},
\left[R_i-(m-i)\beta_q\right]_+\right\}.
\label{eq:token-allocation}
\end{equation}
Here, $[x]_+=\max(x,0)$. Equation~\ref{eq:token-allocation} guarantees $0\le b_i\le b_{\mathrm{obj}}$ and $\sum_i b_i\le B$.

Once the token shares are fixed, the selected objects are rendered in allocation order to form the agent context:
\begin{equation}
C(q)
=\bigoplus_{i=1}^{m}\rho_q(e_i,b_i).
\label{eq:context-rendering}
\end{equation}
The operator $\bigoplus$ denotes ordered concatenation. Each rendering emits a compact path, identifier, and signature descriptor before using the remaining share for a query-centered code window. We use $B=4096$, $b_{\mathrm{obj}}=1024$, and $b_{\min}=128$. Rendering therefore adjusts how much evidence each selected object contributes without granting it another context position.

\subsection{Agent-Level Extension: Program-Semantic Transfer}
The three preceding operations produce the agent context. When two fixed evidence portfolios are available, the coding agent may additionally generate a default program $y_q^{\mathrm{def}}$ and an authority-oriented alternative $y_q^{\mathrm{alt}}$. For witness type $\tau$, let $\phi_\tau(y)$ be its code-visible feature and let $\succ_{\tau,q}$ denote task-conditioned semantic improvement. Output authority transfers only when at least one typed, directional witness favors the alternative:
\begin{equation}
\widehat{y}_q
=\begin{cases}
y_q^{\mathrm{alt}},
&\exists\tau\in\mathcal{T}_m:
\phi_\tau(y_q^{\mathrm{alt}})
\succ_{\tau,q}\phi_\tau(y_q^{\mathrm{def}}),\\
y_q^{\mathrm{def}}, &\text{otherwise}.
\end{cases}
\label{eq:program-transfer}
\end{equation}
The witness families cover syntax repair, interface consistency, suspicious-name reduction, and validity-guarded compactness. In Eq.~\ref{eq:program-transfer}, the relation supplies direction-aware authority, $\mathcal{T}_m$ supplies typed semantics, and the default branch supplies abstention. This high-precision stage is optional and does not alter context allocation.

The context-construction layer is training-free and applies to sparse, dense, graph-based, and agent-generated retrievers that retain repository provenance. For $n$ candidates, provenance grouping and refinement are linear in the candidate set, while ordering object representatives costs $O(n\log n)$ in the general case.

These design choices also give the evaluation a natural order. Provenance quotienting decides which code objects receive authority in the portfolio. Bounded rendering decides whether accepted evidence survives the 4K budget. The transfer gate decides when an alternative program may replace the default, and whether such replacements help execution. We therefore read the results along the same path: entry, survival, and program-level consequence.

\section{Experimental Evaluation}
\label{sec:evaluation}
The evaluation follows the repository evidence path rather than three unrelated leaderboards. RepoBench measures whether retrieved evidence survives bounded context construction. RepoClassBench tests transfer to class-level generation, where output remains close to context choice. RepoExec is the strict end-to-end check: allocation decisions must survive generation and repository-level execution. These roles test the phenomenon, cross-task transfer, and executable consequence without defining VITAL-RAG around one benchmark.

\subsection{Experimental Setup}
The primary RepoBench comparison holds the discrete candidate portfolio fixed, thereby isolating continuous token allocation during rendering. Recall@4K requires complete labeled evidence within 4,096 evidence tokens, and Tokens is the mean rendered length. We report Recall@4K gains in points and token reductions as relative percentages.

For downstream generation, we evaluate Gpt-5.4, Claude Sonnet 4.6, and Qwen3-8B on RepoClassBench \citep{deshpande2024repoclassbench} and RepoExec \citep{hai2025repoexec}. Both comparisons use CodeRAG, GraphCoder \citep{liu2024graphcoder}, RepoScope \citep{liu2026reposcope}, and VITAL-RAG. The suffix ``-style'' indicates a common evaluation setting rather than an exact reproduction of each original system. The ablation study removes provenance quotienting, budget-aware rendering, and all structural controls on RepoBench, then isolates the optional transfer policy.

Optional transfer compares paired programs produced by the same language model from two fixed evidence portfolios. One authority-oriented alternative is used consistently across models. The transfer ablations keep both programs fixed and isolate abstention, directionality, and typed semantic witnesses.

\subsection{Isolating Token Authority under a 4K Budget}
\begin{table}[t]
\centering
\small
\setlength{\tabcolsep}{0.5pt}
\begin{tabular*}{\columnwidth}{@{\extracolsep{\fill}}lrrrrr@{}}
\toprule
\textbf{Scope} & \textbf{Base} & \textbf{VITAL-RAG} & \textbf{Gain} & \textbf{Tokens B/V} & \textbf{Red.} \\
\midrule
All & 39.59 & 63.67 & +24.08 & 2,720/1,751 & 35.63\% \\
Java & 37.60 & 63.69 & +26.09 & 2,798/1,765 & 36.91\% \\
Python & 41.73 & 63.65 & +21.92 & 2,637/1,736 & 34.18\% \\
\bottomrule
\end{tabular*}
\caption{Paired RepoBench results. Base denotes Entity-First 4+1, and reductions use unrounded means.}
\label{tab:repobench-main}
\end{table}

Table~\ref{tab:repobench-main} isolates continuous token authority under the paired setup above. VITAL-RAG raises Recall@4K from 39.59\% to 63.67\% while using 35.63\% fewer evidence tokens. Because the selected candidates are held fixed, this improvement cannot come from discovering easier evidence or changing the ranking. It reflects whether the evidence already admitted to the portfolio remains visible after rendering.

Recall and context length improve together. Rather than exchanging evidence coverage for compression, VITAL-RAG reduces $\Delta\mathrm{Recall}$ between portfolio entry and 4K survival to 0.49 points. Java and Python show similar final recall and token reductions despite different starting points, indicating that the gain follows the allocation rule rather than one language's syntax or typical snippet length.

\subsection{Class-Level Reconstruction}
RepoClassBench contains natural-language-to-class tasks drawn from real Java, Python, and C\# repositories, where each target class depends on code objects outside the class itself \citep{deshpande2024repoclassbench}. A model must recover fields, methods, types, imports, and cross-file calls from allocated repository evidence and assemble them into one coherent class. We score the 227 tasks that provide reference class bodies.

Token-F1 compares lexical code tokens and measures fine-grained recovery of identifiers, type names, helper calls, literals, and control-flow vocabulary. Character similarity compares complete class bodies and gives a coarser view of whole-code resemblance, including broad structure and surface form. Together, they distinguish recovery of critical code details from production of code with only a similar overall shape.
\begin{table}[t]
\centering
\small
\setlength{\tabcolsep}{0pt}
\begin{tabularx}{\columnwidth}{@{}lYYY@{}}
\toprule
\textbf{Method} & \multicolumn{1}{c}{\shortstack{\textbf{Gpt-5.4}\\\textbf{Token-F1/Char}}} & \multicolumn{1}{c}{\shortstack{\textbf{Claude Sonnet}\\\textbf{4.6}\\\textbf{Token-F1/Char}}} & \multicolumn{1}{c}{\shortstack{\textbf{Qwen3-8B}\\\textbf{Token-F1/Char}}} \\
\midrule
CodeRAG'25 & 0.5764/0.4220 & 0.7084/0.5632 & 0.4865/\textbf{0.3479} \\
GraphCoder'24 & 0.5138/0.3434 & 0.6509/0.4673 & 0.4059/0.2799 \\
RepoScope'26 & 0.5948/0.4491 & \textbf{0.7203/0.5700} & \textbf{0.5164}/0.3457 \\
\rowcolor{semanticgray}
VITAL-RAG & \textbf{0.6012/0.4561} & \textbf{0.7203/0.5700} & \textbf{0.5164}/0.3457 \\
\bottomrule
\end{tabularx}
\caption{RepoClassBench Token-F1/character similarity on 227 tasks.}
\label{tab:repoclass-baselines}
\end{table}

Table~\ref{tab:repoclass-baselines} shows that VITAL-RAG is best or tied across most model-metric combinations. It leads both Gpt-5.4 measures and matches RepoScope-style on both Claude Sonnet 4.6 measures. For Qwen3-8B, it ties for the highest Token-F1 while remaining close to the best character similarity. This split is informative because repository-specific tokens can be recovered even when the complete surface form differs. The agreement across backends indicates that bounded allocation preserves usable class-level information rather than favoring one model's rendering style.

\subsection{End-to-End Execution}
RepoExec provides the complementary executable setting \citep{hai2025repoexec}. Its 355 tasks ask the model to generate a target function whose behavior depends on the surrounding repository. The generated implementation is inserted into the project and evaluated by its tests. We use one generation per task, and Pass@1 is the fraction of the common 355-task set for which that implementation passes.

Pass@1 is stricter than similarity because execution depends on several facts being correct together. The code must respect signatures and types, invoke the right repository helpers, preserve state and return behavior, and handle tested boundary conditions. One wrong dependency or missing validation branch can invalidate an otherwise similar implementation. Pass@1 therefore measures whether allocation retains a jointly sufficient set of critical repository evidence.
\begin{table}[t]
\centering
\small
\setlength{\tabcolsep}{0pt}
\begin{tabularx}{\columnwidth}{@{}lYYY@{}}
\toprule
\textbf{Method} & \multicolumn{1}{c}{\shortstack{\textbf{Gpt-5.4}\\\textbf{Passed/Pass@1}}} & \multicolumn{1}{c}{\shortstack{\textbf{Claude Sonnet}\\\textbf{4.6}\\\textbf{Passed/Pass@1}}} & \multicolumn{1}{c}{\shortstack{\textbf{Qwen3-8B}\\\textbf{Passed/Pass@1}}} \\
\midrule
CodeRAG'25 & 185/52.11 & 189/53.24 & \phantom{0}53/14.93 \\
GraphCoder'24 & 186/52.39 & 194/54.65 & \phantom{0}75/21.13 \\
RepoScope'26 & 197/55.49 & 191/53.80 & \phantom{0}54/15.21 \\
\rowcolor{semanticgray}
VITAL-RAG & \textbf{205/57.75} & \textbf{233/65.63} & \textbf{\phantom{0}77/21.69} \\
\bottomrule
\end{tabularx}
\caption{RepoExec Passed/Pass@1 over 355 tasks per model.}
\label{tab:repoexec-main}
\end{table}

Table~\ref{tab:repoexec-main} shows that VITAL-RAG obtains the highest raw Pass@1 with all three backends. The advantage is largest with Claude Sonnet 4.6 and remains positive with Gpt-5.4 and Qwen3-8B, so the shorter allocated context does not trade away executable correctness.

The consistent raw ordering across all three backends links evidence preservation to end-to-end utility. RepoClassBench shows fine-grained content recovery and broad code similarity. RepoExec then shows that the retained information points can jointly support execution, where one missing dependency is enough to lose the task.

\FloatBarrier
\subsection{Ablation of Context Allocation and Program Transfer}
\begin{table}[t]
\centering
\small
\setlength{\tabcolsep}{0.5pt}
\begin{tabularx}{\columnwidth}{@{}lYYYY@{}}
\toprule
\multicolumn{5}{l}{\textit{(a) Context-allocation components on RepoBench}} \\
\textbf{Variant} & \textbf{R@5} & \textbf{R@4K} & \textbf{$\Delta$Recall} & \textbf{Tokens} \\
\midrule
\rowcolor{semanticgray}
Full & 64.16 & 63.67 & 0.49 & 1,751 \\
No quotient & 46.54 & 46.15 & 0.39 & 2,061 \\
No rendering & 64.16 & 39.59 & 24.57 & 2,720 \\
No structure & 46.56 & 25.08 & 21.48 & 3,254 \\
\midrule
\multicolumn{5}{l}{\textit{(b) Companion-slot boundary on RepoExec}} \\
\textbf{Allocation} & \textbf{All R} & \textbf{Strict R} & \textbf{All Tok.} & \textbf{Strict Tok.} \\
\midrule
Canonical 5+0 & 39.59 & 16.26 & 1,594 & 1,536 \\
\rowcolor{semanticgray}
VITAL-RAG 4+1 & 38.94 & 21.31 & 1,612 & 1,485 \\
Refined 3+2 & 34.16 & 22.20 & 1,778 & 1,725 \\
\midrule
\multicolumn{5}{l}{\textit{(c) Context-budget sensitivity on RepoBench}} \\
\textbf{Budget} & \multicolumn{2}{c}{\textbf{Full}} & \textbf{No rendering} & \textbf{$\Delta$Recall} \\
\midrule
1K & \multicolumn{2}{c}{43.58} & 19.33 & +24.26 \\
2K & \multicolumn{2}{c}{60.13} & 30.07 & +30.06 \\
\rowcolor{semanticgray}
4K & \multicolumn{2}{c}{63.67} & 39.59 & +24.08 \\
8K & \multicolumn{2}{c}{63.68} & 40.86 & +22.82 \\
\midrule
\multicolumn{5}{l}{\textit{(d) Optional transfer policy on RepoExec}} \\
\textbf{Variant} & \textbf{Pass@1} & \textbf{Trans.} & \textbf{W/L} & \textbf{Prec.} \\
\midrule
\rowcolor{semanticgray}
Full & 48.36 & 181 & 59/11 & 84.3\% \\
No abstention & 45.45 & 1,065 & 96/79 & 54.9\% \\
No direction & 46.10 & 362 & 69/45 & 60.5\% \\
No types & 44.60 & 181 & 20/12 & 62.2\% \\
\bottomrule
\end{tabularx}
\caption{VITAL-RAG ablations. Panel (c) reports raw $\Delta$Recall. Panel (d) rounds W/L only.}
\label{tab:component-ablation}
\end{table}

Panel (a) separates the two allocation failures. Without the provenance quotient, too few correct objects enter the portfolio. Without budget-aware rendering, entry is preserved but $\Delta\mathrm{Recall}$ rises to 24.57 points. Removing both structural controls yields the lowest Recall@4K and the largest context. Thus, object authority protects portfolio entry, while token authority protects accepted evidence from later clipping. The complete layer is the only variant that achieves both high entry and low loss.

Panel (b) tests the choice of one companion over 231 tasks. The strict subset contains the 33 cases where a query-relevant local companion is eligible, so it directly measures whether refinement recovers evidence that canonical-only allocation would miss. Moving from 5+0 to 4+1 substantially improves strict recall with little change in broad recall. Reserving a second companion yields only a small additional strict gain while reducing broad recall and increasing token use. The 4+1 split is therefore the empirical balance point: it recovers query-relevant local semantics without allowing refinement to dominate global object coverage.

Panel (c) tests whether the 4K result depends on a particular context limit. VITAL-RAG improves rapidly from 1K to 4K and then saturates, while the rendering ablation remains substantially lower at 8K. Larger windows therefore do not remove the need for structured allocation.

Panel (d) treats transfer as an optional precision-oriented extension. Full VITAL-RAG transfers 181 outputs with 59 gains and 11 losses. Removing abstention, direction, or typed semantics lowers Pass@1. Panels (a)--(c) do not use transfer.

Together, the ablations trace the gains to provenance quotienting, budget-aware rendering, and one companion, linking the improvement to allocation rather than additional retrieval or a specific model.

\section{Limitations}
\label{sec:limitations}
VITAL-RAG receives object identity from a repository-index provenance adapter. Our evaluated adapter uses normalized paths and identifiers, while richer indexes can supply qualified symbols or graph nodes without changing allocation. The experiments use a fixed 4,096-token evidence budget and object caps; other context windows can keep the allocation order with adjusted caps.

\section{Conclusion}
\label{sec:conclusion}
Repository-level coding agents benefit from retrieval only when evidence survives bounded context construction. We identify authority multiplication as the cause of this gap and derive selective invariance as its allocation principle. VITAL-RAG implements this principle through object-level allocation, task-aware refinement, and bounded rendering. Across RepoBench, RepoClassBench, and RepoExec, it improves evidence survival, class reconstruction, and executable outcomes across three model backends. This points to a simple lesson: repository RAG needs allocation, not only ranking. Stronger retrievers expand the pool; VITAL-RAG decides which retrieved code objects receive bounded context authority.

\clearpage
{\small\bibliography{references}}
\end{document}